\documentclass[fleqn,preprint,12pt,number]{elsarticle}
\usepackage{amssymb}
\usepackage{hyperref}
\usepackage{placeins}
\usepackage{listings}
\usepackage{textcomp}
\usepackage{lineno}

\journal{Journal of Computational Physics}

\begin{document}

\begin{frontmatter}



\title{A new set of efficient SMP-parallel 2D Fourier subroutines}


\author[UNM,Landau]{Alexander~O.~Korotkevich}

\address[UNM]{Department of Mathematics \& Statistics, The University of New Mexico,
MSC01 1115, 1 University of New Mexico, Albuquerque, New Mexico, 87131-0001, USA}
\address[Landau]{L.\,D.~Landau Institute for Theoretical Physics,
2 Kosygin Str., Moscow, 119334, Russian Federation}

\begin{abstract}
Extensive set of tests on different platforms indicated that there is a performance drop of current standard {\sl de facto} software library
for the Discrete Fourier Transform (DFT) in case of large 2D array sizes (larger than $16384\times16384$).
Parallel performance for Symmetric Multi Processor (SMP)
systems was seriously affected. The remedy for this problem was proposed and implemented as a software library for 2D
out of place complex to complex DFTs. Proposed library was thoroughly tested on different available architectures and hardware
configurations and demonstrated significant ($38-94\%$) performance boost on vast majority of them. The new library together
with the testing suite and results of all tests is published as a project on GitHub.com platform under free software license (GNU GPL v3).
Comprehensive description of programming interface as well as provided testing programs is given.
\end{abstract}

\begin{graphicalabstract}
\includegraphics[width=10.0cm]{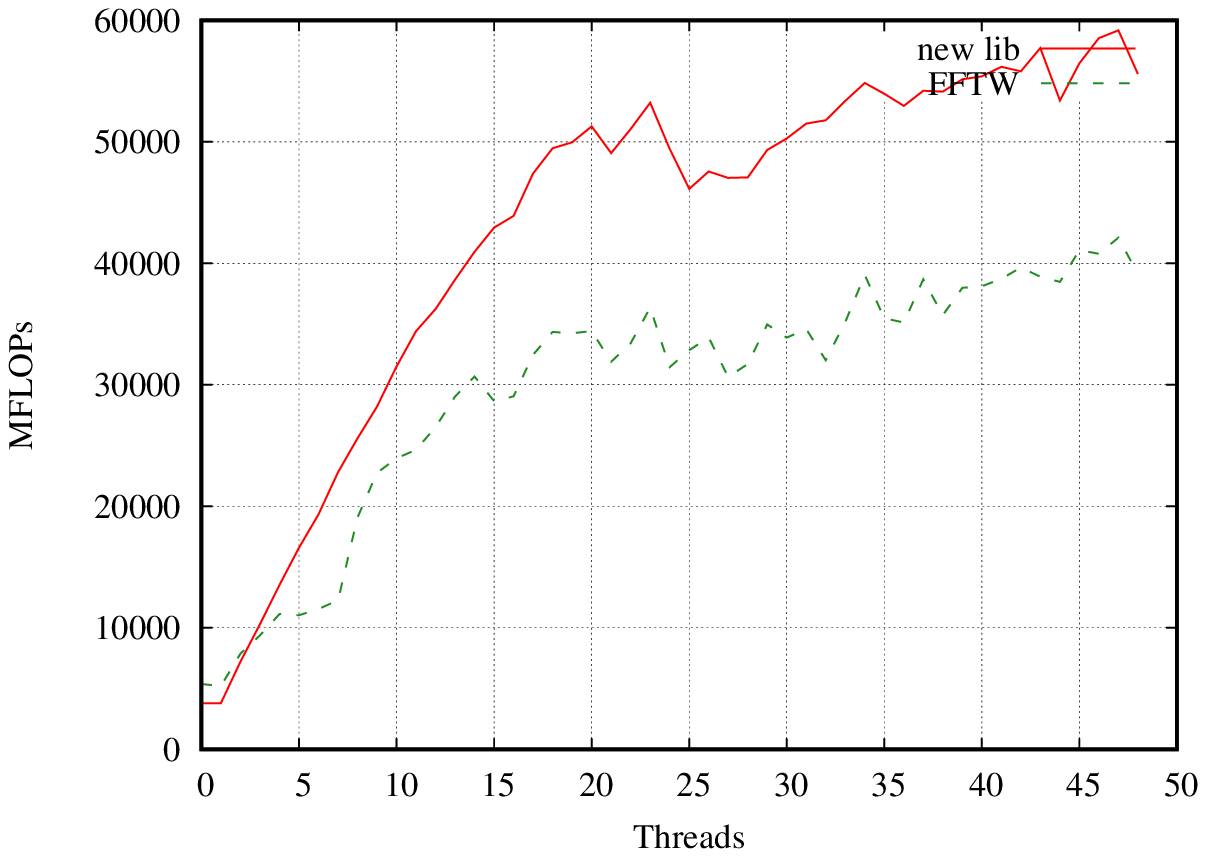}
\end{graphicalabstract}

\begin{highlights}
\item New set of efficient SMP-parallel DFTs
\item Free software license
\item Significant performance boost for large sizes of arrays.
\end{highlights}


\end{frontmatter}


\section{Introduction\label{Intro}}
Since 1990s the standard {\sl de facto} for scientific high performance computations involving discrete Fourier transform (DFT) is FFTW library~\cite{FFTW}.
Author used it during his whole research career, from early 2000s, when it was still version 2. This library provides self tuned perfected version of fast
Fourier transform algorithm, which was initially introduced for broad use in~\cite{CT1965}. From the very beginning of the usage of the FFTW,
author was fascinated by the elegant design of the software and convenient application programming interface (API). Author used FFTW for many
research codes(e.g.~\cite{DKZ2003cap,DKZ2003grav,DKZ2004,Korotkevich2008PRL,KDZ2016,KL2011,KLR2015,DLK2013,DLK2016}).
Some of them involved 2D FFTs, but for very
modest sizes of arrays. For quite a long time to get a computational workstation with many computing central processing unit (CPU) cores was prohibitively
expensive, while distributed memory computations for these sizes of arrays was utterly useless due to communication cost.
Fortunately, situation changed drastically. Now even several tens of CPU cores SMP configurations are readily available for a manageable cost.
Available amount of memory also have risen to hundreds of gibibytes. It made some long standing problems from Physics and Applied Mathematics approachable.
For example,
turbulence simulations require wide range of scales, in other words large wavenumbers arrays, like in the recent work~\cite{FV2015}. In order to check
some theoretical predictions for propagation of the laser radiation in a turbulent atmosphere, like in paper~\cite{KLL2020}, one needs 2D array
of at least (!) $(43000)^2$ size. Taking into account previously described availability of systems with high number of CPU cores,
usage of shared memory program
architecture makes algorithms both simpler to implement and more efficient, as it allows to avoid communication between parallel parts of the program due to
direct access to the whole program memory by any part (thread) of the code. FFTW has multithreaded parallel functions for all kinds of DFTs during
practically all time of its existence. At the same time, when performance tests were done, author was disappointed to notice drastic drop of performance
of multithreaded DFTs with growth of the array sizes (see Figures~\ref{Thor.size_test}-\ref{Old.size_test} below).
It should be specified, that author was mostly interested in 2D DFTs. While prallelization of 2D DFT looks obvious
and one would expect linear or near to linear growth of performance with the number of parallel threads, the problem is not that simple for small
transformation sizes, as inevitable overhead for threads initialization and synchronization can eliminate all the gain if amount of work for every parallel
part is not large enough. But for large, tens of thousands points in every direction 2D arrays situation supposed to be pretty simple and speedup
from parallelization looks inevitable. On the contrary, parallel FFTW performance degradation was increasing with the size of the arrays (as it is
demonstrated later in the paper). A remedy for the situation was found, nearly linear speedup in number of used CPU cores was achieved.
A set of 2D DFTs (out of place, double precision complex numbers) is published~\cite{GitHub_my_fft} under free software license (GNU GPL v3) for use by
community and/or incorporation into existing software.

\section{Proposed solution\label{Solution}}
Author decided that most probably architecture of FFTW is tuned for efficient memory handling for relatively small transformations. This is why
it was decided to use 1D FFTW subroutines and efficient CPU cache memory handling techniques in order to try to improve the situation. For one of
the previous codes~\cite{KL2011} author developed a parallel version of block matrix transposition, which allowed to avoid unnecessary penalties
for cache memory misses (when accessed memory parts are not going one after another, which results in access to the parts of memory (RAM) not mapped into
CPUs cache memory, causing latency delays). As a result, the 2D DFT was implemented in the following way: parallel 1D DFTs (using 1D DFTs from FFTW)
in the direction of array which goes ``along the memory'' (in C programming language matrix organization is so called ``row major format'', which means
that matrix is stored in memory as row after row, unlike FORTRAN's ``column major format'' for a matrix storage, so we first compute 1D DFTs along rows)
are computed; then array is transposed, resulting in other dimension being along the memory (in C columns become rows); now parallel DFTs are computed
for now fast dimension (in C it will correspond to 1D DFTs along rows of transposed matrix, meaning columns of original one);
then we transpose array again. Operations of array transposition look
unnecessary, as one could just perform 1D DFTs along second dimension without them, but in such a case distance in memory between successive elements
of corresponding 1D array will be large (for large arrays) and nearly every operation with these elements would result in CPU ``cache miss'',
meaning the CPU will wait for data from RAM instead of number crunching. As it was mentioned above we
avoid unnecessary ``cache misses'' by using block array transposition. For large array sizes (when array doesn't fit into CPU cache memory)
effect of efficient memory handling could easily results in order of magnitude speedup of transposition operation in comparison with naive version of it.
Block transposition is a pretty standard technique demonstrating importance of thoughtful use of memory especially for large array sizes (some discussion
of even more advanced memory handling techniques used internally in FFTW can be found, for example, here~\cite{CacheOblivious}). Simple set of subroutines
for pool of threads organization and synchronization was used~\cite{GitHub_my_thr} (included with the new library). Because author created this set of 2D DFT subroutines specifically for his set
of research tasks, only complex-to-complex out-of-place DFTs were implemented. Other types of DFTs could be implemented by demand or if new research needs
will arise. Published code was not specifically optimized for speed or memory usage, with exception of memory block size for transposition subroutine,
which was tested on several architectures to provide slight performance gain with respect to some generic value
(only two different values: for AMD\textregistered\, and Intel\textregistered\, CPUs). Application programming interface (API) was made as close to FFTW's one as possible. Is described in
details in~\ref{AppendixAPI}, together with complete description of benchmark program and proposed performance test procedure.

\section{Performance of the code\label{Performance}}
All systems which participated in benchmark are briefly described in \ref{AppendixSystemsDescr}.
In all the tests version 3.3.8 (the latest at the moment of writing the article) of FFTW was used. Configuration of the library for different CPUs and machines
is described in \ref{AppendixFFTWConf}.
For evaluation of performance the same methodology
as described for official FFTW benchmark utility (described in~\cite{FFTW_bench}) was used, which means that for complex transforms number of operations
is estimated as $5N\log_2 N$ (here $N$ is the total number of harmonics/points),
time (up to nanoseconds) is measured for a series of DFTs which takes at least several (usually tens or hundreds)
of seconds to compute, then this procedure is repeated $10$ times and smallest time (meaning the best performance) expressed in microseconds was
used to divide estimated number of operations in order to get performance in MFLOPs (Mega-FLOPs or millions of floating point operations per second).
Two series of tests were performed: fixed array size and variable number of parallel threads, fixed number of threads (usually the number of threads
which gives the best performance for FFTW in the previous test was chosen) and variable size of the array in order to determine at what value performance drop happens.

\subsection{Fixed array size. Performance scaling with number of threads\label{thr_test}}
\begin{figure}[htbp]
\centering
\includegraphics[width=10.0cm]{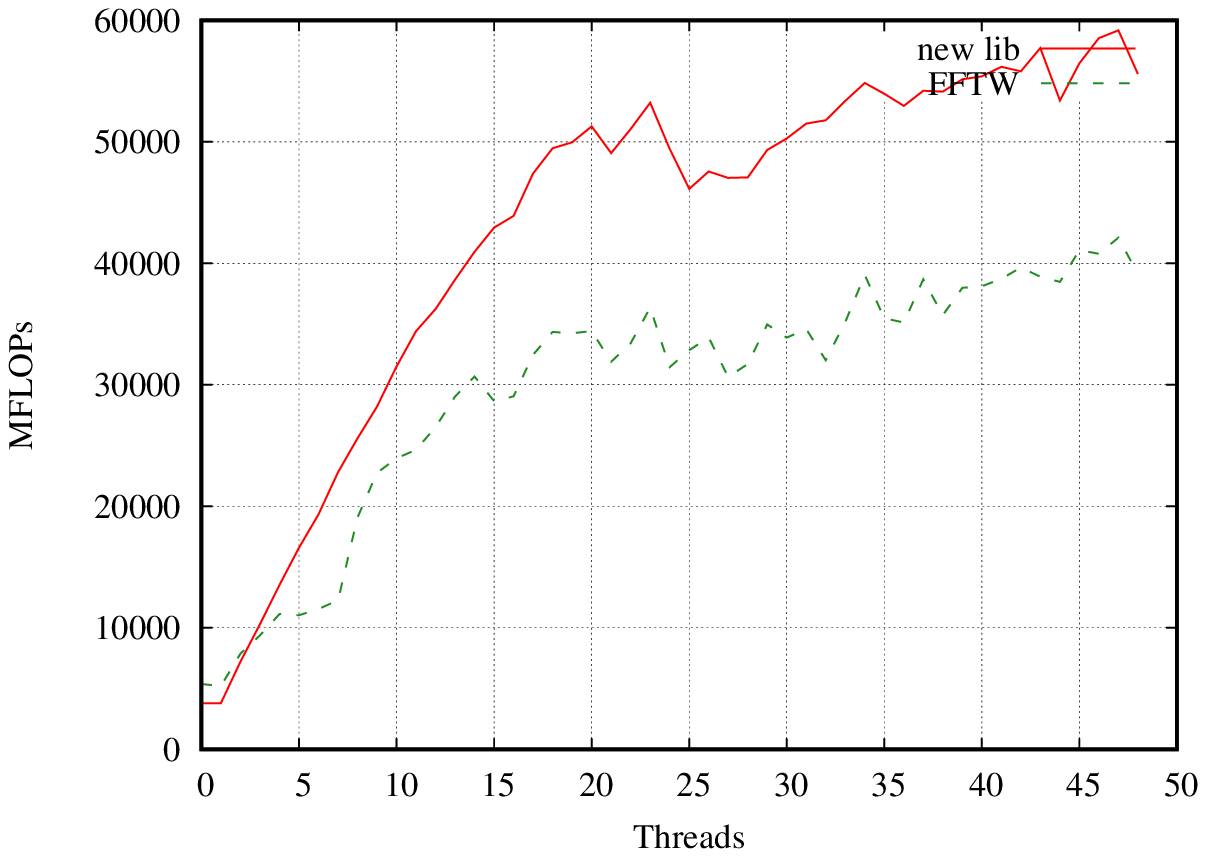}
\caption{\label{Thor.thr_test} Array size $(32768)^2$, performance dependence as a function of number of parallel threads. Red: new library, green: FFTW. {\bf System 1} was equipped with two
Intel\textregistered\, Xeon\textregistered\, CPU E5-2680 v3 @ 2.50GHz (Haswell), which results in $2\times12$ CPU cores, $2\times24$ HyperThreading cores.}
\end{figure}

\begin{figure}[htbp]
\centering
\includegraphics[width=6.78cm]{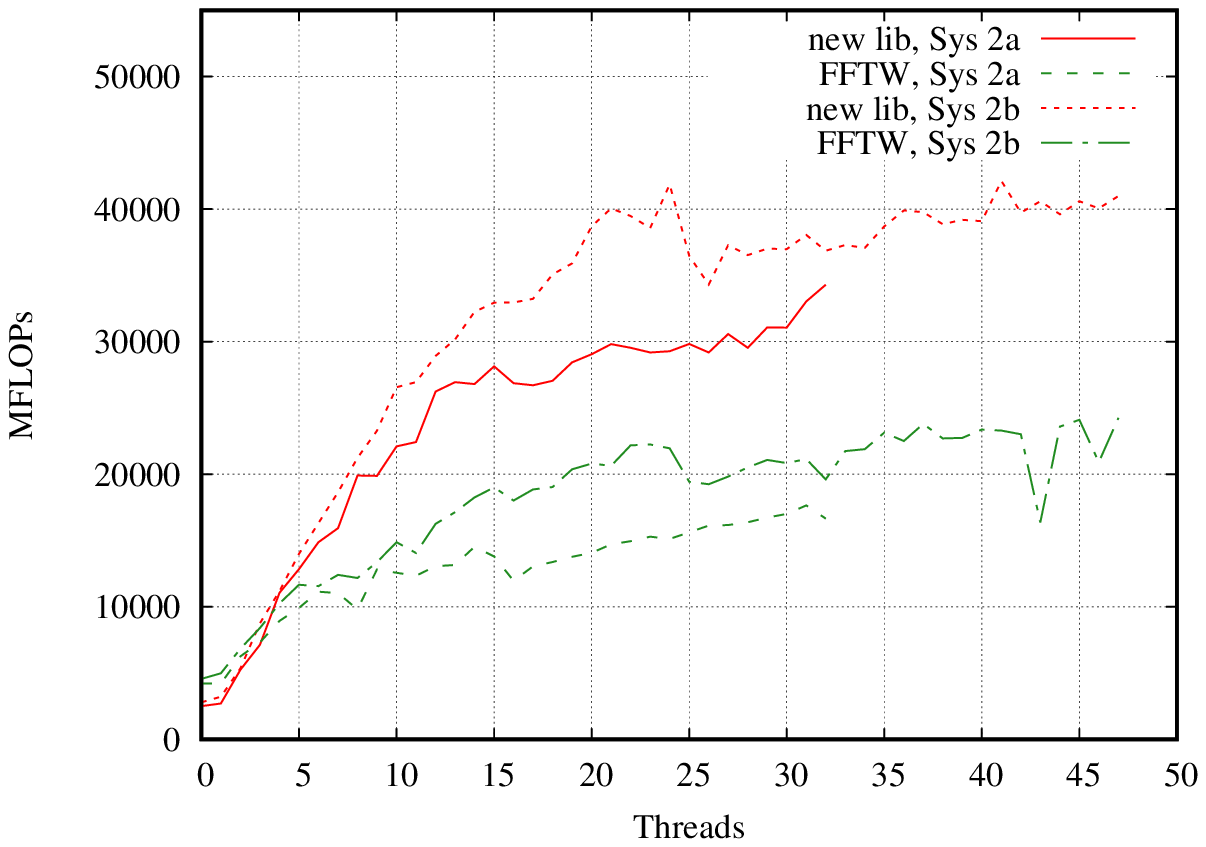}
\includegraphics[width=6.78cm]{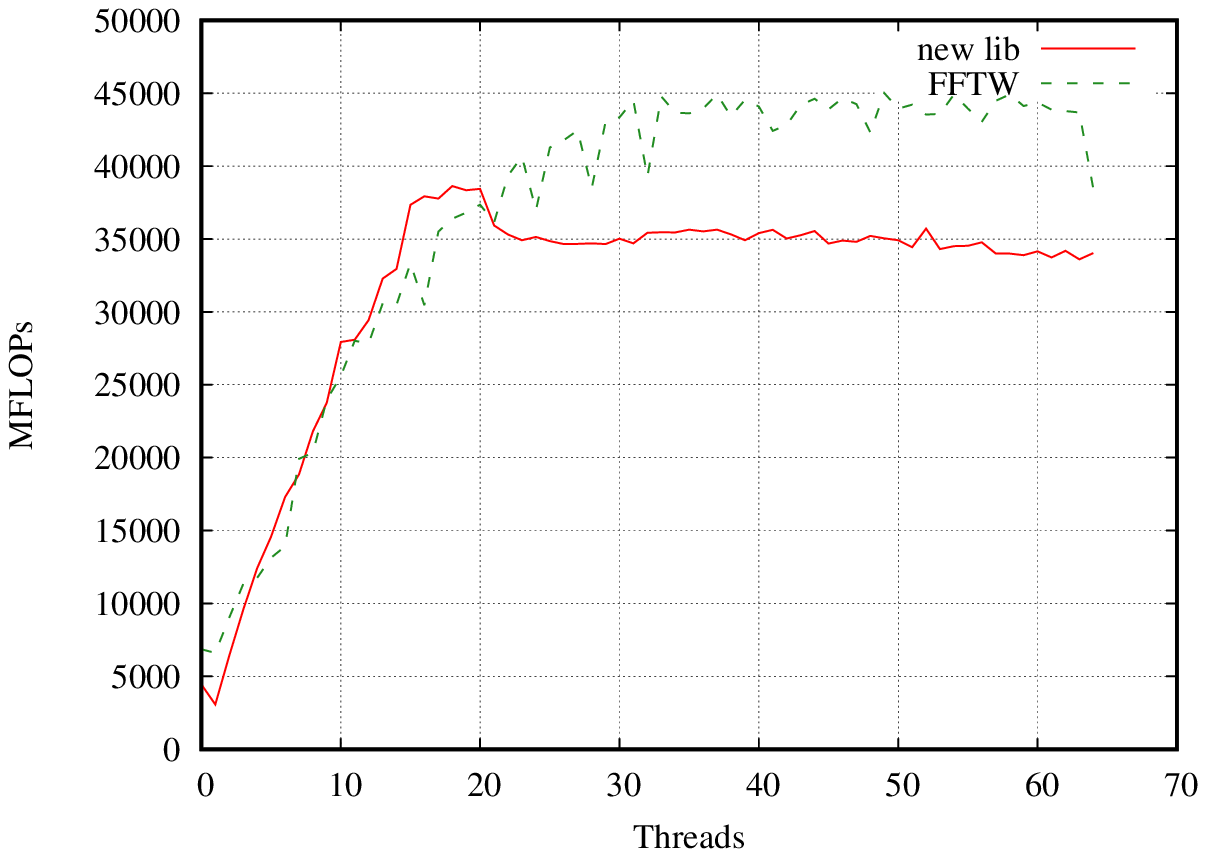}
\caption{\label{New.thr_test} Array size $(32768)^2$, performance dependence as a function of number of parallel threads. Red: new library, green: FFTW. {\sl (Left panel)} {\bf System 2a} was equipped
with two Intel\textregistered\, Xeon\textregistered\, Silver 4110 CPU @ 2.10GHz (Skylake),  which results in $2\times8$ CPU cores, $2\times16$ HyperThreading cores. {\bf System 2b} was equipped with two Intel\textregistered\, Xeon\textregistered\, Gold 6126 CPU @ 2.60GHz (Skylake), which results in $2\times12$ CPU cores, $2\times24$ HyperThreading cores. {\sl (Right panel)}
{\bf System 3} was equipped with two Intel\textregistered\, Xeon\textregistered\, Gold 6242 CPU @ 2.80GHz (Cascade Lake), which results in $2\times16$ CPU cores, $2\times32$ HyperThreading cores.}
\end{figure}

\begin{figure}[htbp]
\centering
\includegraphics[width=6.78cm]{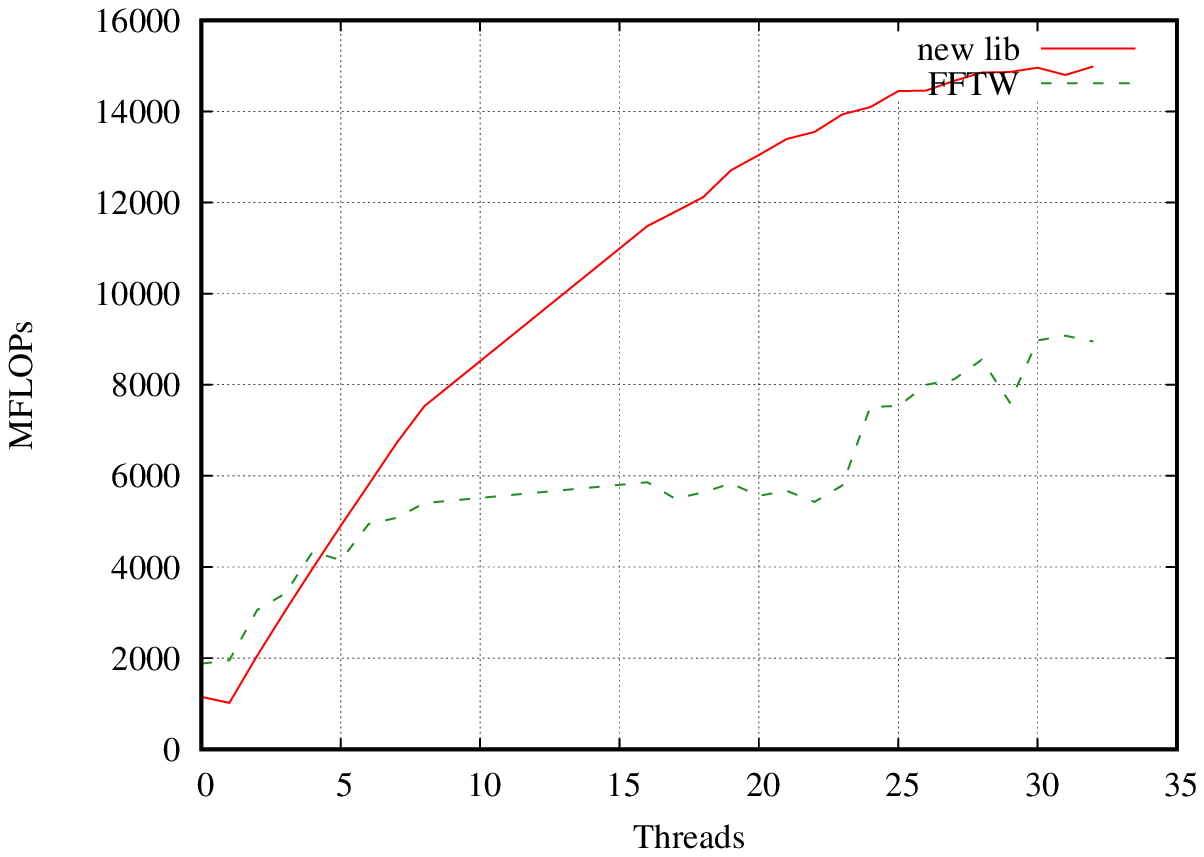}
\includegraphics[width=6.78cm]{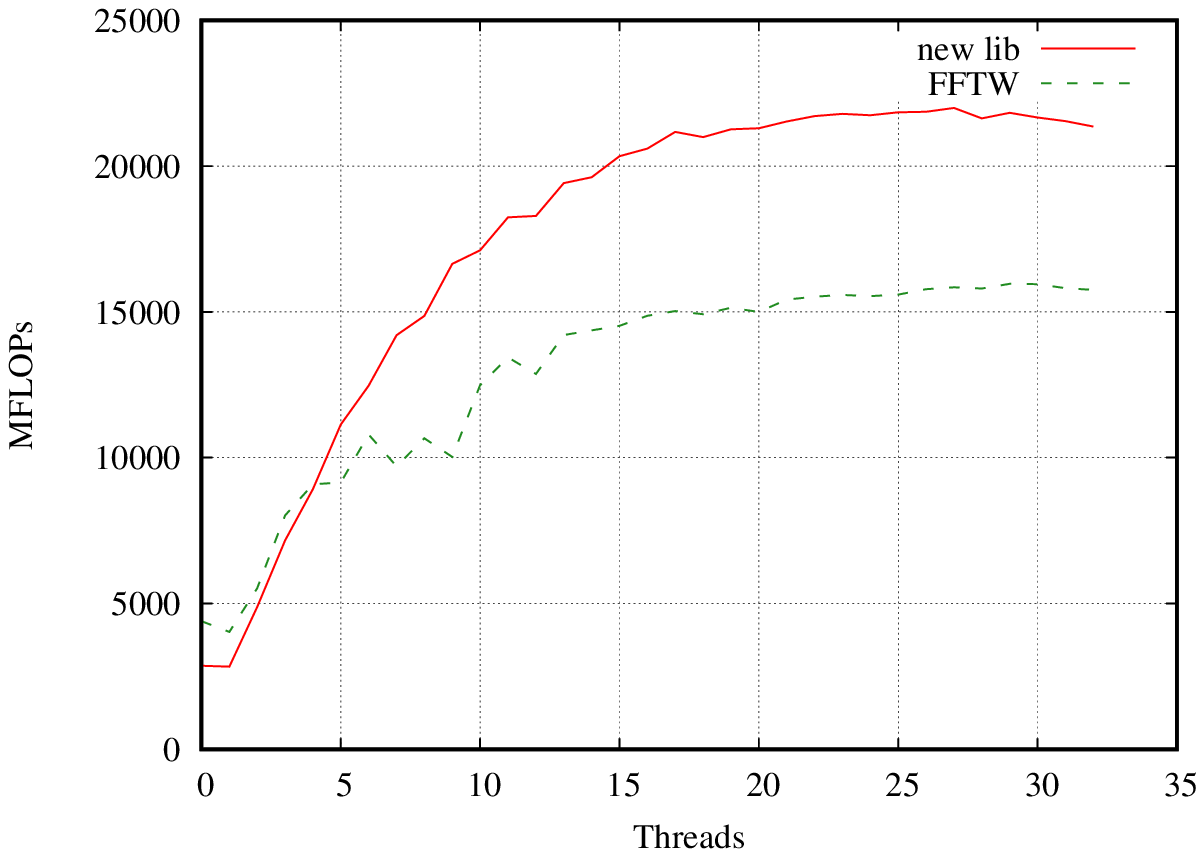}
\caption{\label{Old.thr_test} Array size $(32768)^2$, performance dependence as a function of number of parallel threads. Red: new library, green: FFTW. {\sl (Left panel)} {\bf System 4} was equipped
with two AMD\textregistered\, Opteron\texttrademark\, Processor 6276 @ 2.3GHz (Bulldozer), which results in $2\times8$ CPU cores, $2\times16$ HyperThreading cores. {\sl (Right panel)}
{\bf System 5} was equipped with two Intel\textregistered\, Xeon\textregistered\, CPU E5-2670 0 @ 2.60GHz (Sandy Bridge EP), which results in $2\times8$ CPU cores, $2\times16$ HyperThreading cores.}
\end{figure}

In all the cases double precision complex array of the size $32768\times32768$ was used. Out of place (means different input and output arrays are used)
complex to complex DFTs were evaluated for both new library and FFTW. Benchmark program is included in the library package.
Its usage together with methodology of the tests are explained in the end of~\ref{AppendixAPI}.
For all but one system (exception was {\bf System 2b}) the full test was completed
twice with several days between runs to avoid influence of system processes etc. Best results of all runs were used for evaluation. Benchmark results are represented
in Figures~\ref{Thor.thr_test}-\ref{Old.thr_test}.


As one can see, for this size of the array new library gives performance boost for all but one system
(exception is {\bf System 3}). See Table~\ref{tab:thr_test} for a summary of results.
\begin{table}[htb!]
\center
\begin{tabular}{|c|c|c|c|c|c|c|}
\hline
MFLOPS & {\bf Sys 1} & {\bf Sys 2a} & {\bf Sys 2b} & {\bf Sys 3} & {\bf Sys 4} & {\bf Sys 5}  \\
\hline
FFTW & $42109$ & $17640$ & $24241$ & $44918$ & $9078$ & $15961$  \\
\hline
new library & $59175$ & $34292$ & $42130$ & $38628$ & $14986$ & $21993$  \\
\hline
boost   & $41\%$ & $94\%$ & $74\%$ & $-14\%$ & $65\%$ & $38\%$  \\
\hline
best/core/GHz & $986$ & $1021$ & $675$ & $501$ & $407$ & $529$  \\
\hline
\end{tabular}
\caption{\label{tab:thr_test} Comparison of best performances for a problem $32768\times32768$ double precision complex array. Numbers for FFTW
and new library are given in MFLOPS. It should be noted that these numbers could be achieved at different numbers of used parallel threads for FFTW and new library. ``Boost'' shows speedup of the best results of the new library with respect to the best result of FFTW. Also we provide estimation of efficiency
for different systems, computed as best performance (regardless of the library) divided by number of real CPU (not HyperThreading) cores and by base frequency in GHz.}
\end{table}
Performance improvement ranges for most systems from 38\% to 94\%. The only system which behaved completely differently (see
Figure~\ref{New.thr_test}, right panel) is {\bf System 3}. Although architectures of {\bf Systems 2a,b} and {\bf System 3} are very close,
the dependence changed drastically after some number of parallel threads. As one can see, until number of threads close to $16$ everything looks
pretty similar to other systems. Some ideas why it can be so are given in the end of~\ref{AppendixSystemsDescr}. Also, one could notice
that $16$ is exactly the number of real CPU cores on one of the two processors. As a result, absence of speedup after adding more parallel threads
can be attributed to some problems of communication between two CPUs in the system. The source(-s) of these problems can range from
architecture flaws and/or hardware configuration to configuration of Linux kernel or even motherboard (BIOS) software. The fact that there is no
noticeable influence of this issue on FFTW's performance could be the lower usage of memory throughput due to slightly lower performance which does not
create some kind of issue in the kernel and/or CPUs. As a summary, the behavior of {\bf System 3}
is still a mystery for the author. Even on this system if we limit number of parallel threads by $16$ (one real CPU)
new library demonstrates improvement in performance.
\FloatBarrier

\subsection{Fixed number of threads. Performance scaling with array size\label{size_test}}
\begin{figure}[htbp]
\centering
\includegraphics[width=10.0cm]{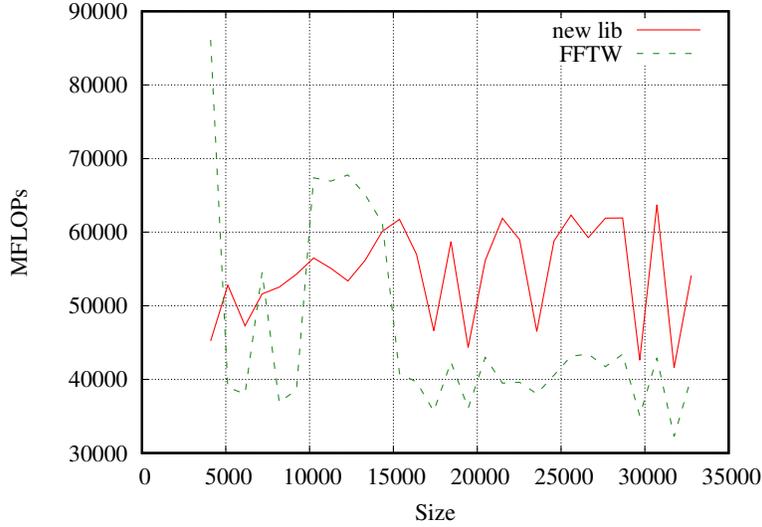}
\caption{\label{Thor.size_test} Performance dependence as a function of array size $N$, full 2D-array size if $N^2$. Number of threads was chosen as giving the best performance to FFTW in the previous series of tests. Red: new library, green: FFTW. {\bf System 1} was equipped with two
Intel\textregistered\, Xeon\textregistered\, CPU E5-2680 v3 @ 2.50GHz (Haswell), which results in $2\times12$ CPU cores, $2\times24$ HyperThreading cores, number of threads is $47$.}
\end{figure}

\begin{figure}[htbp]
\centering
\includegraphics[width=6.78cm]{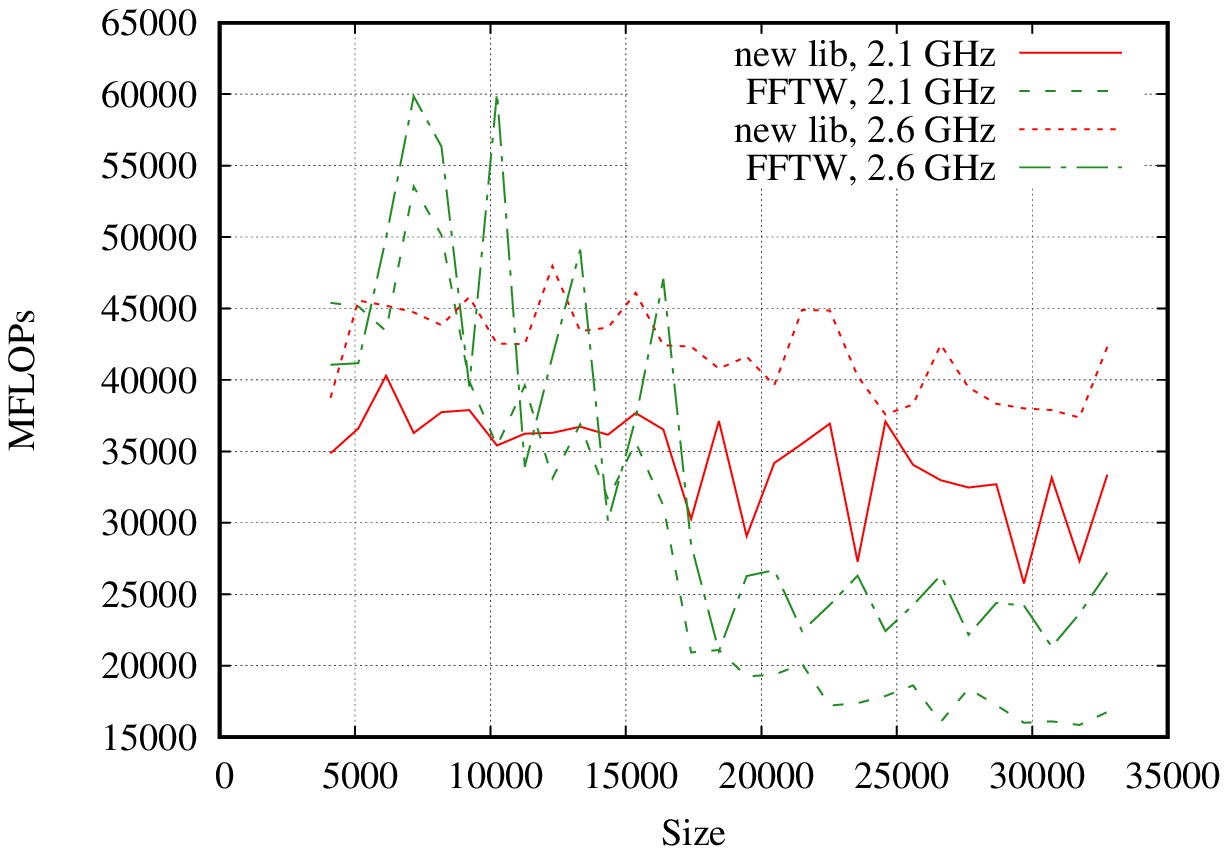}
\includegraphics[width=6.78cm]{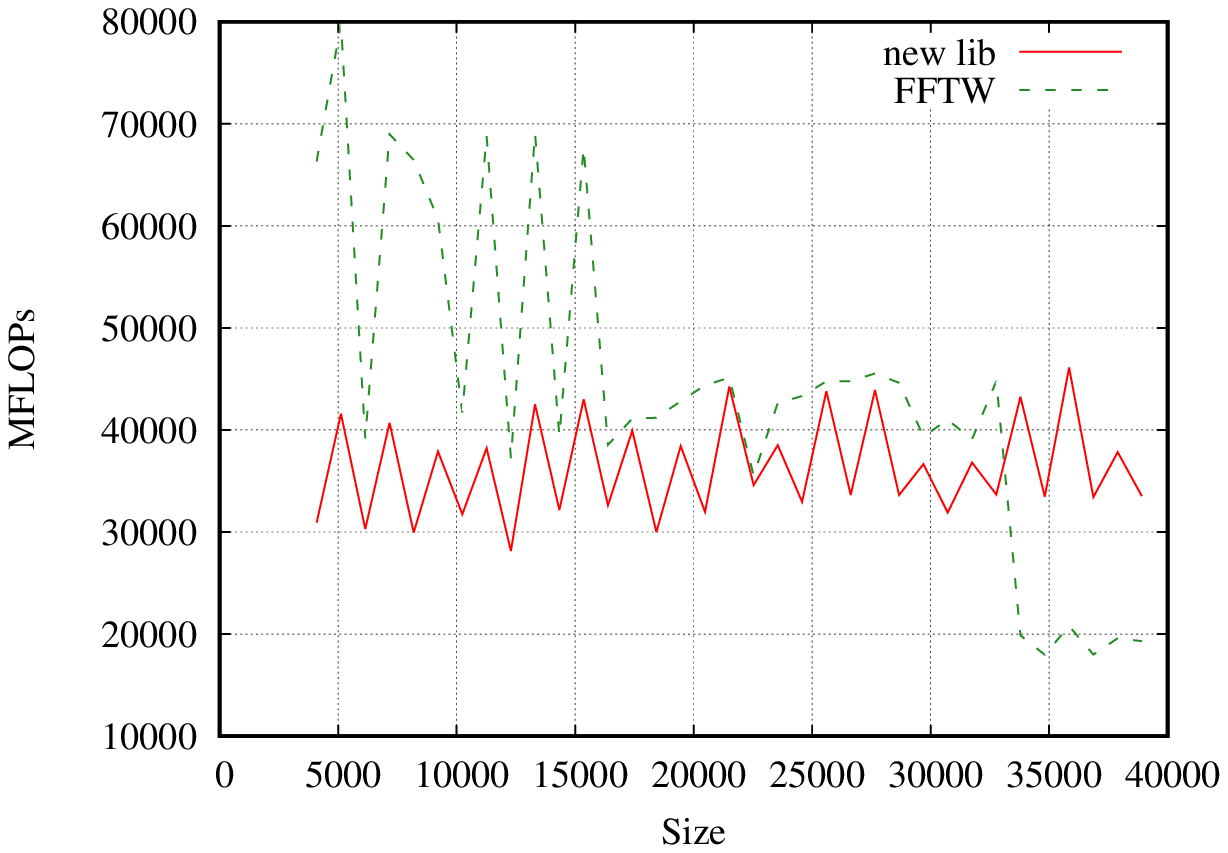}
\caption{\label{New.size_test} Fixed number of threads (giving best performance to FFTW in the previous series of tests), performance dependence as a function of size (length of the side) of the square array. Red: new library, green: FFTW. {\sl (Left panel)} {\bf System 2a} was equipped
with two Intel\textregistered\, Xeon\textregistered\, Silver 4110 CPU @ 2.10GHz (Skylake),  which results in $2\times8$ CPU cores, $2\times16$ HyperThreading cores, number of threads is $32$. {\bf System 2b} was equipped with two Intel\textregistered\, Xeon\textregistered\, Gold 6126 CPU @ 2.60GHz (Skylake), which results in $2\times12$ CPU cores, $2\times24$ HyperThreading cores, number of threads is $48$. {\sl (Right panel)}
{\bf System 3} was equipped with two Intel\textregistered\, Xeon\textregistered\, Gold 6242 CPU @ 2.80GHz (Cascade Lake), which results in $2\times16$ CPU cores, $2\times32$ HyperThreading cores, Number of threads is $33$.}
\end{figure}

\begin{figure}[htbp]
\centering
\includegraphics[width=6.78cm]{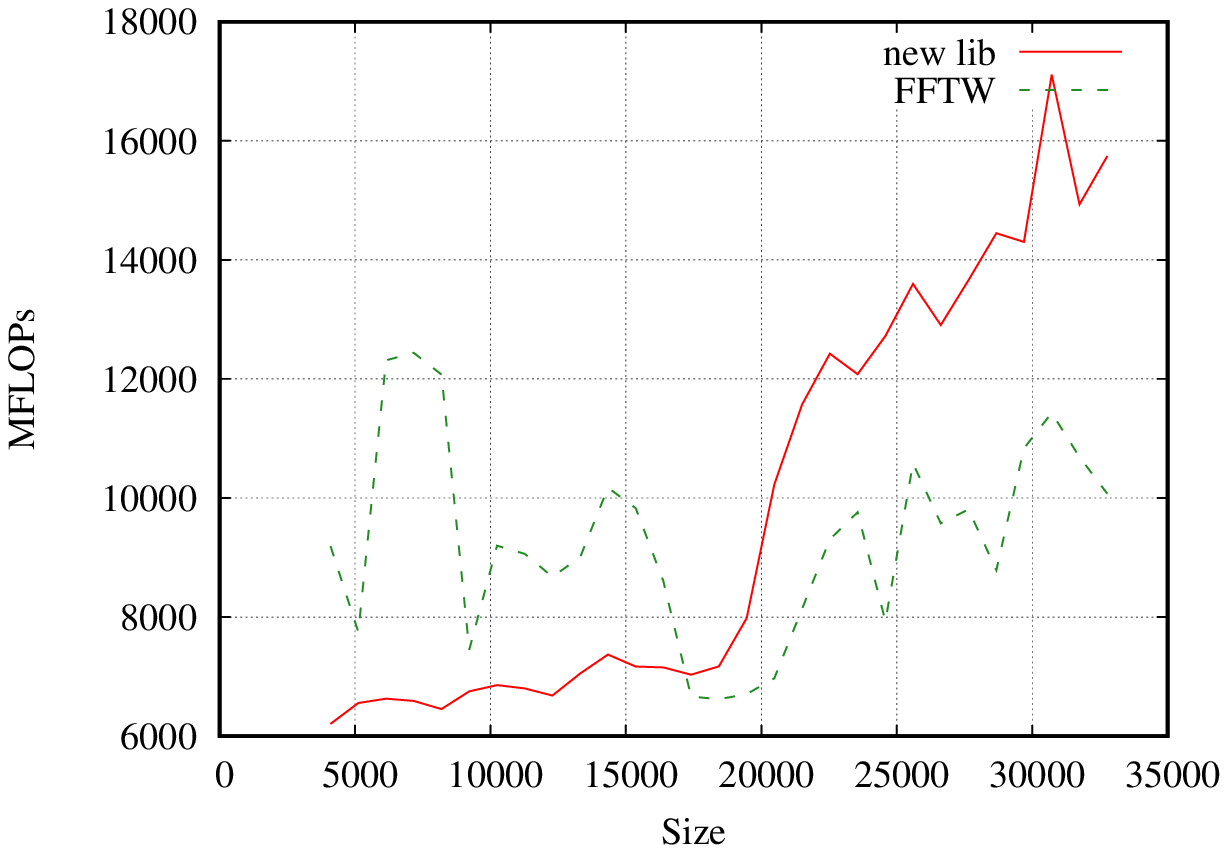}
\includegraphics[width=6.78cm]{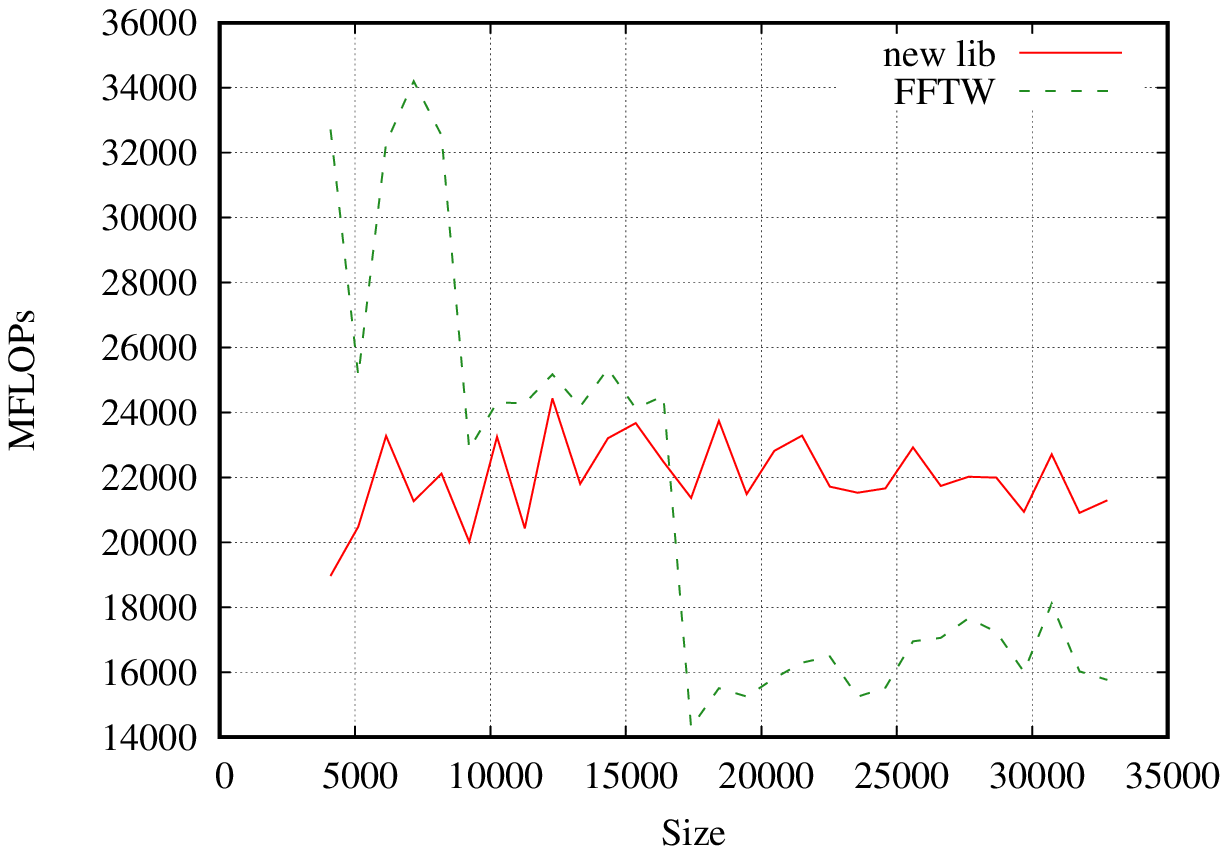}
\caption{\label{Old.size_test} Fixed number of threads (giving the best performance to FFTW in the previous series of tests), performance dependence as a function of size (length of the side) of the square array.. Red: new library, green: FFTW. {\sl (Left panel)} {\bf System 4} was equipped
with two AMD\textregistered\, Opteron\texttrademark\, Processor 6276 @ 2.3GHz (Bulldozer), which results in $2\times8$ CPU cores, $2\times16$ HyperThreading cores, number of threads is $32$. {\sl (Right panel)}
{\bf System 5} is equipped with two Intel\textregistered\, Xeon\textregistered\, CPU E5-2670 0 @ 2.60GHz (Sandy Bridge EP), which results in $2\times8$ CPU cores, $2\times16$ HyperThreading cores, Number of threads is $32$.}
\end{figure}

In order to test the performance of two libraries under consideration as a function of the size of an array, author performed another series of tests.
Now the number of parallel threads was fixed and only the size of the problem was changing. Once again, only out of place double precision complex
DFTs were tested. Number of threads was determined from the tests described in the previous Subsection~\ref{thr_test}. Namely, author used number
of parallel threads which gave the highest (or close to that) performance for FFTW library on every architecture. All arrays were square and
the problem size was starting from $4096\times4096$ with continuation in integer (natural) multiples of $1024$ (for example, $(20\times1024)^2$).
The same testing program included in the library package was used. Results of the tests are represented in Figures~\ref{Thor.size_test}-\ref{Old.size_test}.


As can be seen from the measurements, for every architecture the size of the array when FFTW experiences a dramatic drop in performance is
slightly different.
At the same time the performance of the new library relatively weakly depends on the size of the problem. Observed oscillations are the consequence of
different efficiency of 1D FFTW DFTs for different sizes of the problems (``out of the box'' FFTW  is the most efficient for sizes which can be
represented as products of integer (natural) powers of few lowest prime numbers $2$, $3$, $5$, and $7$, although there is a possibility to make FFTW to be as
efficient for larger primes as well~\cite{FFTW_sizes}). As it can be noted, dips and peaks of the performance of new library perfectly coincide
with ones for FFTW, which is inevitable as the same 1D DFT subroutines are used. See Table~\ref{tab:size_test} for a summary of results.
\begin{table}[htb!]
\center
\begin{tabular}{|c|c|c|c|c|c|c|}
\hline
Array sizes & {\bf Sys 1} & {\bf Sys 2a} & {\bf Sys 2b} & {\bf Sys 3} & {\bf Sys 4} & {\bf Sys 5}  \\
\hline
Threshold & $16\times2^{10}$ & $17\times2^{10}$ & $18\times2^{10}$ & $16\times2^{10}$ & $17\times2^{10}$ & $17\times2^{10}$  \\
 &  &  &  & $33\times2^{10}$ &  &   \\
\hline
Perf. drop   & $35\%$ & $40\%$ & $45\%$ & $25\%$ & $30\%$ & $40\%$  \\
   &  &  &  & $50\%$ &  &   \\
\hline
\end{tabular}
\caption{\label{tab:size_test} Dependence of FFTW performance on problem's size given as $S\times1024=S\times2^{10}$, where $S$ is an integer (natural) number,
starting from $4\times1024$. Here by threshold size author means one side of the square array, e.g. size $17\times1024$ means that the actual problem
was $17408\times17408$, double precision complex array. Performance drops are only approximate, as DFT performance oscillates with the size of the problem.}
\end{table}
In all the cases there is a significant $25-40\%$ performance degradation when size of the array exceeds the threshold close to $16384\times16384$
(slightly different for different systems).
Again we see different behavior of {\bf System 3} with respect to other systems. The first drop in performance by $25\%$ happens, as problem size exceeds
$16\times1024$, the second one, twice stronger ($50\%$ if we are using relative measure, but approximately the same amount in MFLOPS),
after approximately double of the first threshold, i.e. at $33\times1024$. Taking this fact into account, one could see, that if the previous performance
tests were done not for array's size $32\times1024$, but for $33\times1024$, perhaps we would get result similar to other systems.
\FloatBarrier

\section{Conclusion}
The significant degradation of performance of current standard {\sl de facto} DFT library, using multiple parallel threads on
SMP systems, motivated the author to perform
extensive set of tests on different platforms. Results of these tests indicated that there is performance drop
for array sizes exceeding some threshold (slightly different for different CPUs and architectures). The author implemented a relatively simple
algorithm as a library using 1D DFT subroutine from FFTW together with parallel block transposition subroutine. The new library demonstrated
significant ($35-94\%$) performance boost on all but one configurations. Even in the exceptional case slight increase in performance was achieved
for some limited number of parallel computational threads. The new library together
with the testing suite and results of all tests are published as a project on GitHub platform under free software license (GNU GPL v3).
Comprehensive description of programming interface as well as provided testing programs is given in the~\ref{AppendixAPI}.

Author would like to mention that domination of Intel\textregistered\, CPUs in the list of tested systems was not intentional.
AMD\textregistered\, Opteron\texttrademark\, CPU of Bulldozer family (and similar) are infamous for their dreadful performance. Newer
CPUs of the same company appeared few years ago. Author tried to contact AMD\textregistered\, through different means of communication
and requested a remote access to any SMP machine for 24 hours. Author's request was declined, it was proposed to contact vendors. Unfortunately,
during the time of tests no high performance computing vendors offered servers equipped with recent AMD\textregistered\, CPUs.

Current version of the library provides only 2D out of place complex to complex DFT subroutines. It comes with some limitations and not optimized
for memory consumption. Some straightforward ways of improvement include different pools of threads for transposition and 1D DFTs,
elimination of current limitations on dimensions of the arrays (currently they have to be multiples of relatively small number, like $64$),
and optimization of memory usage. Further development will depend on
author's future research needs as well as demand from potential users. Author would prefer the new library to be included as a possible
engine for large arrays into existing libraries and will try to work in this direction.
Another option will be internal tuning of the existing libraries, perhaps based on tests presented in this paper, in order to eliminate
existing significant performance degradation for large arrays. Any of these two variants will make life of users (including the author)
way easier as there will be no need to choose different libraries for different sizes of the problems.

\section*{Acknowledgments}
The author has performed this work for the project supported by the Simons Collaboration on Wave Turbulence and author is grateful for this support.
Some of the systems used for development and testing ({\bf System 1} and {\bf System 4}) were funded by
NSF grant OCE 1131791. Author would like to express his gratitude to A.\,O.~Prokofiev, who explained him
block transposition in early 2000s.
Also author would like to thank developers of FFTW~\cite{FFTW} for this beautiful and free example of
software engineering.

\appendix
\section{Library description\label{AppendixAPI}}
\subsection{Current library packaging}
As currently there is only one subroutine for out of place 2D complex to complex DFT, author left the library as just mere C-files, which
have to be compiled into object files and linked to the users program in the usual way. As the interest level to the library is not yet clear, author
would prefer to invest the time into creation of auto-configuration and installation script only if requested by other users.

The new library is dependent on FFTW v3. Author strongly recommends to compile users local version of the library
in the user's directory, configured to use system specific SIMD instructions as explained in~\ref{AppendixFFTWConf}, as performance gain can be
substantial. After that {\tt Makefile} has to be edited in order to provide paths to header files and library files. Currently, static linking is used.
If user prefers to use dynamic linking the option {\tt -static} in {\tt LDFLAGS} has to be erased. Usual command {\tt make clean all} will
compile the library (object files) and the benchmark program.

\subsection{Library API}
Library API was intentionally made as close to FFTW's one as possible. In the description we suppose that the reader is familiar with FFTW's API, detailed
description of which can be found at FFTW's web-site~\cite{FFTW}. First of all, one needs to initialize threads (here we initialize 32 threads):
\begin{verbatim}
unsigned long int threads_number=32;

my_fft_init (threads_number);
\end{verbatim}
Here {\tt threads\_number} is the number of threads to be initialized in the pool of threads. Maximum number of threads to be available for the computations.
This value has to be more or equal to zero. 

Then the usual creation of a DFT's plan has to be performed:
\begin{lstlisting}[breaklines]
my_fft_plan plan_fwd;

plan_fwd = my_fft_plan_dft_2d (in, out, out2, NX, NY, +1, FFTW_EXHAUSTIVE, threads_number);
\end{lstlisting}
Here {\tt in} and {\tt out} are input and output arrays similarly to FFTW's API. Array {\tt out2} is a scratch array of the same size which is necessary
for the present version as the array transposition subroutine does not support in place functionality. This limitation can be easily eliminated
in the next version. All other parameters are passed directly to the 1D DFT API of the FFTW library. Here {\tt threads\_number} is actual number of threads
to be used for DFT (has to be less or equal to number of threads previously specified in the {\tt my\_fft\_init}). If it is equal to zero, nonparallel
(linear) version of subroutine is used.

{\bf Important note:} Currently library supports only sizes when both {\tt NX} and {\tt NY} are multiples of variable {\tt BLOCK\_SIDE\_SIZE}
defined in the file {\tt my\_fft\_lib.h}. This variable defines a size of the square block used for transposition. This limitation can be
eliminated if needed. Even in its present form it hardly pose any practical issue, as sizes of arrays when this library become
beneficial are usually $16\times1024$ or larger
for {\tt NX} and {\tt NY}, as a result to change it to the nearest multiple of even $64$ (optimal case for all tested Intel\textregistered\, CPUs, for AMD\textregistered\, it is
equal to $16$) is not a serious problem.

In order to perform planned DFT one needs to call the subroutine {\tt my\_fft\_execute} with the previously created plan:
\begin{verbatim}
my_fft_execute (plan_fwd);
\end{verbatim}

In the case if the user needs to free the resources allocated for the previously created plan, one needs to call the following subroutine:
\begin{verbatim}
my_fft_destroy_plan (plan_fwd);
\end{verbatim}

In order for the functions definitions to be included into the {\tt *.c}-file one needs to include {\tt my\_fft\_lib.h}:
\begin{verbatim}
#include "my_fft_lib.h"
\end{verbatim}
Here it is supposed that file {\tt my\_fft\_lib.h} is located in the same directory.

If you would like to use FFTW concept of wisdom (information about previously computed optimal plans, significantly accelerates creation
of new plans), just load wisdom from file using standard FFTW subroutine. It should be noted, that new library uses {\bf only} 1D linear DFTs and
parallel and linear wisdom are not compatible. Please, load only usual (nonparallel) wisdom, not the one produced as a result of plans creation for parallel
DFTs.

\subsection{Benchmark program}
The simple benchmark program {\tt my\_fft\_test} is provided together with the library files. It allows to perform a performance test for
specified size of an array using both new library's and FFTW's subroutines. Only out of place complex to complex 2D DFTs are compared.
Result is reported in MFLOPS. Here is an example of usage of the program:
\begin{lstlisting}[breaklines]
$ ./my_fft_test 
Usage is the following:
./my_fft_test NX NY num_of_repeats number_of_threads [output_file]
$ ./my_fft_test 8192 8192 10 8
NX = 8192, NY = 8192, num_reps = 10, threads_number=8
8 Threads initialized.
Wisdom successfully imported.
Exporting wisdom to file...
Time for 10 Fourier transforms using my_fft is 1.09193526530e+01 seconds.
........ (here will be output for 10 runs)
Time for 10 Fourier transforms using my_fft is 1.11323441190e+01 seconds.
MFLOPS = 8062.030694
Using FFTW_MEASURE option for FFTW plan. Usually quite fast.
Time for 10 Fourier transforms using FFTW is 9.55567851900e+00 seconds.
........ (here will be output for 10 runs).
Time for 10 Fourier transforms using FFTW is 9.55813624100e+00 seconds.
MFLOPS = 9256.224182
Normed L_infty norm: ||Mismatch between my_fft and fftw||_infty/(NX*NY) = 5.218373960228812e-17.
\end{lstlisting}
Input parameters are: sizes of the array in $X$ and $Y$ dimensions, number of repeated DFTs for every measurement of time (for good results should
give at least few seconds per measurement, several tens of seconds even better), number of threads to use, optional output file name. If the output
file name is provided, result is written to a file (a line is appended to the end of file) as seven columns: $X$-size, $Y$-size, number of threads used,
perfomance in MFLOPS for new library, perfomance in MFLOPS for FFTW, best measured time for new library, best measured time for FFTW. Also
shell-scripts for testing are provided. They will be discussed in the next subsection together with methodology of testing.

{\bf Important note 1:} Author used {\tt FFTW\_MEASURE} option for creation of FFTW parallel plans for large arrays, although FFTW benchmarking
manual~\cite{FFTW_bench} recommends to use {\tt FFTW\_PATIENT}. The reason is {\bf extremely} long time necessary for creation of plans with this
option. For example, on {\bf System 1} creation of a plan for problem size $32768\times32768$ using $47$ threads with {\tt FFTW\_PATIENT} option
took around $30$ hours. No significant advantage {\bf for large array sizes} was obtained. On {\bf System 2b} creation of a plan for problem size $32768\times32768$
using $24$ threads using {\tt FFTW\_PATIENT} option took more than $24$ hours. Again, no significant performance boost {\bf for large array sizes} was noticed. If one would
like to try it, variable {\tt FFTW\_PATIENT\_PLAN} has to be defined in the file {\tt my\_fft\_test.c}.

{\bf Important note 2:} As it was mentioned above, $10$ time measurement runs are performed and the best (smallest) time is used for evaluation
of performance. The number of runs can be changed by redefining variable {\tt RUNS\_NUMBER} in the file {\tt my\_fft\_test.c}.

Program tries to load FFTW wisdom from the file {\tt fftw.wisdom} located in the same directory. If the file is found, the program reads
the wisdom and after creation of a plan for the new library writes the cumulative wisdom to the same file. If program cannot find {\tt fftw.wisdom}
file, it creates plans without it and saves the accumulated wisdom to {\tt fftw.wisdom} file if possible. Unfortunately, as it was already mentioned above,
for FFTW v3 wisdom information for linear and parallel planning subroutines are incompatible. Even more, if linear planning
subroutine was used in the program, parallel wisdom cannot be used in the same program. If this situation will change in future releases,
all necessary parts of the code can be initiated by defining variables {\tt READ\_THR\_WISDOM\_FOR\_FFTW} and {\tt SAVE\_THR\_WISDOM\_FOR\_FFTW}
in the file {\tt my\_fft\_test.c}. For parallel wisdom program will try to use file {\tt fftw.thr.wisdom} located in the same directory.

\subsection{Installation and testing methodology}
When benchmark program is compiled and working properly, one could try to fine tune the only changeable parameter in the library, namely
the variable {\tt BLOCK\_SIDE\_SIZE} in the file {\tt my\_fft\_lib.h}, which determines the side length of the square block used for transposition.
By author's experience, the optimal values are: $64$ for Intel\textregistered\, CPUs and $16$ for AMD\textregistered\, CPUs (author had an access only to relatively old
AMD\textregistered\, CPUs). Author recommends to use powers of $2$. Of course, after every change of this value, library and test program has to be recompiled.

After that one can start scanning system for performance dependence on number of used threads with fixed array size.
Initially, it is good to determine the largest array which fits into system's RAM memory.
Due to limitation mentioned above author recommends to use multiples of $2^{10}=1024$. Let us say that the maximum size of the array is
$(20\times1024)^2$ and the system has $16$ logical CPU cores (usully it means $8$ real CPU cores and $8$ more HyperThreading cores).
There is a simple shell-script {\tt thr\_test.sh} provided with the library. This is how it can be tuned for the situation described above:
\begin{verbatim}
#!/bin/bash

for i in {0..8}; do
        ./my_fft_test 20480 20480 5 $i 20k.User_thr_test.dat
done

for i in {9..16}; do
        ./my_fft_test 20480 20480 10 $i 20k.User_thr_test.dat
done
\end{verbatim}
Here we use $5$ DFTs for every time measurements for small number of parallel threads and $10$ for presumably faster configurations with
larger number of parallel threads. The output file format is described above in details.

After completing the previous scan, one should find the number of threads which gives the best performance for FFTW. This number of threads
will be used to determine the performance dependence on the size of the problem when number of parallel threads is fixed. For example,
let us suppose that the best performance is achieved by FFTW using $15$ threads. Then one can slightly modify simple script {\tt size\_test.sh}
provided with the library in the following way:
\begin{verbatim}
#!/bin/bash

for i in {4..20}; do
        size=$(echo $i*1024| bc)
        ./my_fft_test $size $size 10 15 User_size_test.15thr.dat
done
\end{verbatim}
Here we scan in integer (natural) multiples of $1024$ as the side length of a square array. Size is changing from $4\times1024$ till $20\times1024$.
As we determined previously that this is the largest size which fits into system's RAM. The output file format is described above.

This is the method which was used in all measurement presented in this paper.

\section{Tested systems\label{AppendixSystemsDescr}}
All systems were working under different flavors of GNU Linux
distributions ranging from Ubuntu and Debian to Gentoo. GNU Compiler Collection (gcc) was used as a C compiler in all the cases.
Here are short hardware descriptions for every system.
\begin{itemize}
\item {\bf System 1}\\
CPU: $2\times$ Intel\textregistered\, Xeon\textregistered\, CPU E5-2680 v3 @ 2.50GHz (Haswell), 2x12 CPU cores, 2x24 HT cores\\
RAM: 128 GiB
\item {\bf System 2a}\\
CPU: $2\times$ Intel\textregistered\, Xeon\textregistered\, Silver 4110 CPU @ 2.10GHz (Skylake), 2x8 CPU cores, 2x16 HT cores\\
RAM: 64 GiB
\item {\bf System 2b}\\
CPU: $2\times$ Intel\textregistered\, Xeon\textregistered\, Gold 6126 CPU @ 2.60GHz (Skylake), 2x12 CPU cores, 2x24 HT cores\\
RAM: 64 GiB
\item {\bf System 3}\\
CPU: $2\times$ Intel\textregistered\, Xeon\textregistered\, Gold 6242 CPU @ 2.80GHz (Cascade Lake), 2x16 CPU cores, 2x32 HT cores\\
RAM: 384 GiB
\item {\bf System 4}\\
CPU: $2\times$ AMD\textregistered\, Opteron\texttrademark\, Processor 6276 CPU @ 2.3GHz (Bulldozer), 2x8 CPU cores, 2x16 HT cores\\
RAM: 64 GiB
\item {\bf System 5}\\
CPU: $2\times$ Intel\textregistered\, Xeon\textregistered\, CPU E5-2670 0 @ 2.60GHz (Sandy Bridge EP), 2x8 CPU cores, 2x16 HT cores\\
RAM: 128 GiB
\end{itemize}
It should be noted, that all systems, except {\bf System 3}, were producing very consistent results in all the tests. The {\bf System 3}
behaved very inconsistently, with significantly different results between benchmarking runs (each run took about 24 hours, difference could be as high
as 10\%). This is the newest
CPU in the tested systems and it was released already after recent cache memory associated vulnerabilities reported for Intel\textregistered\, CPUs.
Perhaps, temporary and/or urgent patches to the cache memory system and Linux kernel were the reason for such a behavior. Author has no other explanation.

\section{Configuration of FFTW\label{AppendixFFTWConf}}
For most of the systems ({\bf Systems 1-3}) participated in benchmarking the following line was used during configuration of FFTW library:
\begin{lstlisting}[breaklines]
$ ./configure --enable-threads --disable-fortran --with-gnu-ld --enable-avx2 --enable-fma --enable-sse2
\end{lstlisting}
Of course, if you need FORTRAN interface, please, remove {\tt --disable-fortran} from the line. As {\bf Systems 4-5} do not support AVX2 SIMD (single
instruction multiple data) instructions, slightly different configuration line was used:
\begin{lstlisting}[breaklines]
$ ./configure --enable-threads --disable-fortran --with-gnu-ld --enable-avx --enable-fma --enable-sse2
\end{lstlisting}
It should be noted, that although {\bf Systems 2a,b-3} support AVX512 SIMD instruction set, it is {\bf not} recommended to use configuration option
{\tt --enable-avx512} as it results in drastic drop in performance. Comparison of different configurations can be performed using {\tt bench} program
supplied with FFTW library. The following command line was used for final testing: {\tt ./bench -onthreads=24 -opatient -s oc32768x32768 }, initial testing
was done on smaller $4096\times4096$ array.

As an another note, author do  {\bf not} recommend to supply any additional {\tt CFLAGS} during compilation. In his experience it resulted only in
performance drop. Default compilation flags are well tuned and (nearly?) optimal.

\bibliographystyle{elsarticle-num} 
\bibliography{surfacewaves}



\end{document}